\documentclass[aps,prl,reprint,nofootinbib]{revtex4-1}

\usepackage{hyperref} 

\usepackage{epsfig}
\usepackage{bm}
\usepackage{amssymb}
\usepackage{mathrsfs}
\usepackage{amsmath}
\usepackage{dsfont}
\usepackage{color}
\usepackage{cancel}
\usepackage{booktabs}
\usepackage{upgreek}

\newcommand{\be}{\begin{equation}}
\newcommand{\ee}{\end{equation}}

\newcommand{\F}{ \mathcal F}
\newcommand{\C}{\mathcal C}
\newcommand{\avg}[1]{\left\langle #1 \right\rangle}

\begin{document}
\title{First-principle simulations of 1+1d quantum field theories at $\theta=\pi$ and spin-chains
}
\author{Tin Sulejmanpasic}
\email{tin.sulejmanpasic@durham.ac.uk}
\author{Daniel G\"oschl}
\email{daniel.goeschl@uni-graz.at}
\author{Christof Gattringer}
\email{christof.gattringer@uni-graz.at }
\affiliation{$^*$Department of Mathematical Sciences, Durham University, DH1 3LE Durham, United Kingdom\\
$^{\dagger,}$$^\ddagger$University of Graz, Institute for Physics\footnote{Member of NAWI Graz.},  A-8010 Graz, Austria }

\begin{abstract}
We present a lattice study of a 2-flavor $U(1)$ gauge-Higgs model quantum field theory with a topological term at $\theta=\pi$. Such studies are 
prohibitively costly in the standard lattice formulation due to the sign-problem. Using a novel discretization of the model, along with an 
exact lattice dualization, we overcome the sign-problem and reliably simulate such systems. Our work provides the first 
ab initio demonstration that the model is in the spin-chain universality class, and demonstrates the power of the new approach to $U(1)$ gauge theories.

\end{abstract}

\maketitle

Quantum field theories with $\theta$-terms are of immense interest, both in high-energy as well as condensed matter physics. The 
$\theta$-angle is an example of a purely quantum deformation, which is inconsequential for the classical motion of the system. Yet the presence of 
the $\theta$-term can dramatically change a quantum system. A textbook example of a $\theta$-term is the motion of a particle on a ring 
in the presence of a magnetic flux. Classically the motion is undisturbed as the magnetic field is zero 
everywhere along the path of the particle. Still the quantum system can feel the magnetic field through the Aharonov-Bohm effect, reshuffling the 
spectrum back onto itself as the magnetic flux is increased to the unit flux quantum. The $\theta$-term in the path-integral description precisely 
corresponds to the magnetic flux, normalized such that $\theta=2\pi$ corresponds to the unit flux quantum. 

In Quantum Chromodynamics the possibility to write a $\theta$-term is the basis of the strong CP problem, which is among the 
most important problems of modern high-energy physics. More relevant for this work is that in the effective description of \emph{antiferromagnetic} 
(AFM) spin-chains  $\theta$-terms may arise due to the Berry phases in the path-integral quantization of spin, 
as first noted by Haldane \cite{Haldane:1982rj}. The observation that integer and half-integer spin chains are 
distinguished in the effective field theory description 
by the value of the $\theta$-parameter, $\theta=0$ and $\theta =\pi$, respectively, is the basis of Haldane's conjecture. 

Haldane's work, along with the integrability of the $S=1/2$ Heisenberg model, the idea of 
non-abelian bosonization \cite{Witten:1983ar} and the theoretical
tractability of the \emph{Wess-Zumino-Witten} (WZW) theories \cite{Knizhnik:1984nr,Gepner:1986wi}, gives a compelling self-consistent 
picture of the $\theta=\pi$ abelian field theory, which we will review below. Yet very little is understood from first principle computer simulations. 
The reason is that the introduction of the $\theta$-term gives rise to a complex weight 
in the conventional formulation of the path-integral, which prohibits efficient Monte-Carlo sampling. Recently we have proposed a 
solution to this sign-problem relevant for such systems. 
The approach relies on the reformulation and dualization of $U(1)$ lattice gauge theory in 2d \cite{Gattringer:2018dlw}, that was
generalized to higher dimensions by two of us in \cite{Sulejmanpasic:2019ytl}. Here we 
apply these ideas to a 2-flavor $U(1)$ gauge-Higgs model, relevant for spin-chains. Using first principle numerical calculations we, 
for the first time, confirm the theoretical picture that
arises indirectly by other reasoning. At the same time the consistency gives credence to our novel approach to $U(1)$ lattice gauge theories, 
which have applications also in higher-dimensional spin systems, most notably to the yet unsettled deconfined quantum criticality in (2+1)d anti-ferromagnets (see, e.g., \cite{Senthil:2004aza,Sandvik:2006fpf,Sandvik:2010ag,Kaul:2011dqx} and references therein). On the other hand our formulation may also have interesting implications for fundamental aspects of electrodynamics on the lattice, including electric magnetic duality, and the possible existence of a continuum limit \cite{Sulejmanpasic:2019ytl} contradicting the general lore. The success of our methods demonstrated here in (1+1)d are an important step towards a better understanding of U(1) gauge theories and their interdisciplinary significance.

\vspace{.3cm}

\noindent{\bf The model and its connection to spin chains: } The model we study is the 2-flavor Abelian gauge-Higgs 
model  described by the Euclidean Lagrangian

\begin{multline}\label{eq:model}
\mathcal L= \frac{1}{4e^2}F_{\mu\nu}F^{\mu\nu}+\frac{i\theta}{2\pi}F_{12}\\+(D_\mu \Phi)^\dagger (D_\mu\Phi)+
m^2 \Phi^\dagger\Phi+\lambda (\Phi^\dagger\Phi)^2\;,
\end{multline}
where $\mu=1,2$ is the space-time index, $\Phi=(\phi_1,\phi_2)^T$ is an $SU(2)$ scalar doublet. $F_{\mu\nu}=\partial_\mu A_\nu-\partial_\nu A_\mu$ and $D_\mu=\partial_\mu+i 
A_\mu$, with $A_\mu$ the $U(1)$ gauge field. We will view the above theory in the spirit of an effective theory, so that the Lagrangian should be 
supplemented with a UV cutoff $\Lambda$. A transformation $\Phi\rightarrow U\Phi$ where $U$ is an $SU(2)$ unitary matrix is a symmetry of the 
Lagrangian. However, the $\mathbb Z_2$ center symmetry of $SU(2)$ acting as $\Phi\rightarrow -\Phi$ is a subgroup of the $U(1)$ gauge 
symmetry $\Phi(x)\rightarrow e^{i\varphi(x)}\Phi(x)$, so the global symmetry is ${SU(2)}/{\mathbb Z_2}\cong SO(3)$ instead. Moreover the system 
has a charge conjugation symmetry $\C$ which takes $\Phi\rightarrow \Phi^*$ and $A_\mu\rightarrow -A_\mu$, when $\theta=0$ or $\pi$, which, 
together with $SO(3)$ forms the symmetry group $O(3)$. 

$U(1)$ gauge theories in 2d are natural candidates for spin-1/2 AFM spin-chains. 
The history of the connection of model \eqref{eq:model} to spin-chains is a long one, with some more recent developments involving anomalies, 
which we briefly review here. 

Consider first the limit\footnote{It is conventional in high-energy literature to label the coupling in the Lagrangian 
as $m^2$, such that when $m^2$ is positive, $m$ is the tree-level mass of the $\Phi$ excitations.} $-m^2\ll \Lambda^2$, such that the classical 
potential is minimized at 
the value $\Phi=\Phi_0= u\sqrt{|m^2|/(2\lambda)} $, where $u$ is a 2-component unit vector. Writing $\Phi(x)=u(x)\sqrt{|m^2|/(2\lambda)}$, the limit 
$-m^2\ll\Lambda^2$ effectively sets $u^\dagger u=1$ and the model reduces to the 
weakly coupled $CP(1)$ nonlinear sigma model (NLS), which is equivalent to the $O(3)$ NLS model\footnote{To be precise, the $CP(1)$ model 
has no kinetic term for the gauge field. However, the kinetic term is irrelevant (i.e., the coupling $e$ is relevant), so its presence is not expected to 
change the behavior. Alternatively, we can think of the model with a nonzero kinetic term as the RG iterated $CP(1)$ model, 
where the kinetic term is generated along the RG flow \cite{Witten:1978bc}.}. 

The opposite limit of the model \eqref{eq:model}, where $m^2$ is large and positive is exactly computable, as the $\Phi$-field is massive and can 
be integrated out. The result is a pure gauge theory at $\theta=\pi$, which is exactly solvable and has a double vacuum degeneracy due to the $
\C$-symmetry breaking. 

On the other hand Haldane has shown that the $SU(2)$ Heisenberg spin-chain in the spin $S$ representation is equivalent to the weakly coupled 
$O(3)$ model in the large $S$ limit \cite{Haldane:1982rj}, where the $\theta$-angle is given by $\theta=2\pi S\bmod 2\pi$ for translationally invariant systems, 
indicating that the integer and half-integer $SU(2)$ spin-chains fall into separate universality classes. 
This  is nicely consistent with the Lieb-Shultz-Mattis (LSM) theorem\footnote{The LSM theorem states that an $SO(3)$ and translationally 
invariant antiferromagnetic spin-chain is either gapless, or breaks translation symmetry 
spontaneously. } \cite{Lieb:1961fr,Affleck:1986pq}, which, along with the integrability of the $S=1/2$ Heisenberg spin-chain, 
gives credence to the conjecture that the $O(3)$ model at $\theta=\pi$ is critical (see also \cite{Shankar:1989ee}). 
This in turn leads to the plausible conjecture that all half-integer Heisenberg spin-chains are critical. 
Furthermore, the $\theta=\pi$ $O(3)$ model, as well as its cousin \eqref{eq:model}, are subject to a variety of anomaly matching constraints 
\cite{Gaiotto:2017yup,Komargodski:2017dmc,Komargodski:2017smk,Sulejmanpasic:2018upi,Tanizaki:2018xto} 
-- which should be viewed as  LSM-like theorems. 
\vspace{.3cm}

\noindent{\bf Lattice theory and duality: } The usual lattice discretization uses $U(1)$ valued phases on links. When considering the 2d theory with 
the $\theta$-term, it is, however, useful to instead define $\mathbb R$-valued gauge fields $A_l$ on links $l$, supplemented by integer-valued 
variables $n_p$ living on the plaquettes $p$, with gauge action 
\be
S_G[A,n]=\sum_p \frac{\beta}{2}(F_p+2\pi n_p)^2+i\theta \sum_p n_p\;,
\ee 
where $\beta=\frac{1}{2e^2}$ is the inverse gauge coupling, and $F_p$ the discretized version of the field strength. Apart from the 
$\theta$-term, the above action is the well-known Villain discretization of $U(1)$ lattice gauge theory \cite{Villain:1974ir}, 
while the $\theta$-term was 
introduced in \cite{Gattringer:2018dlw,Sulejmanpasic:2019ytl}\footnote{The $n_p$ play the role of a discrete 2-form gauge field which allows the 
values of the $A_l$ to be restricted to $[-\pi,\pi]$. Moreover, similar reasoning can be used to make a connection \cite{Sulejmanpasic:2019ytl}
with the geometric definition of the topological charge \cite{Luscher:1981zq}.}. Using Poisson resummation it is possible to replace 
\be
\sum_{n_p\in \mathbb Z} \! e^{-\frac{\beta}{2}\big(F_p+2\pi n_p\big)^2 + \,i\theta n_p } \rightarrow 
\sum_{m_p\in\mathbb Z} \! e^{-\frac{1}{2\beta} \big(m_p+\frac{\theta}{2\pi}\big)^2 +\, i F_p m_p},
\ee
such that the action is now linear in $F_p$. Integrating out the $A_l$ after the appropriate ``partial integration'' imposes the constraint that 
for pure gauge theory $m_p$ is constant on all plaquettes, with a remaining weight that is real and positive.

The matter sector of model \eqref{eq:model} is described by an $SU(2)$ bosonic (Higgs) doublet $\Phi$ on lattice sites $x$, with the action
\begin{eqnarray}
S_{H}[\Phi,A] & \; = \; & \sum_{x}\Big[ M \Phi_x^\dagger \Phi_x  + 
\lambda \big(\Phi_x^\dagger \Phi_x\big)^2 
\nonumber \\
& & \quad - \sum_{\mu=1}^{2} 
\big(\Phi_x^\dagger e^{iA_{x,\mu}} \Phi_{x + \hat \mu} \, + \, c.c.\big) \Big] \; ,
\label{S_Higgs}
\end{eqnarray}
where $M=4+m^2$. In (\ref{S_Higgs}) we denote links $l$ as $(x,\mu)$ and $A_{x,\mu}$ is a gauge field on a link rooted at $x$ in the 
direction $\mu$.
The partition function with the above matter-action can be dualized to a sum over closed $U(1)$ currents described by closed contours $\mathcal C$ built out of lattice links, which couple to the gauge field as 
$e^{i\sum_{l\in\C} A_l}$. After the insertion of such wordlines, $A_l$ can be integrated out, causing $m_p$ to jump at the worldlines by 
the amount of $U(1)$ charge carried by the wordline. If the matter field in question is bosonic, as in (\ref{eq:model}), 
the statistical weight of the configurations is strictly positive, allowing for Monte Carlo simulations in the dual representation. 
Using suitable methods \cite{Mercado:2013yta} we simulate the model \eqref{eq:model},
varying $m^2$ at fixed $\lambda = 0.5$ and $\beta=3$. See \cite{SM} for details.

\vspace{0.3cm}

\noindent{\bf Phase diagram and numerical results:} The LSM theorem states that if $O(2)\subset SO(3)$-spin and lattice translations are good 
symmetries of a half-integer spin chain, then either the spin chain is gapless, or gapped and degenerate. On the other hand, the field theory at $\theta=\pi$ has an analogous `t Hooft anomaly involving the charge-conjugation symmetry $\C$ and the spin symmetry 
$SO(3)$, implying that either the $\C$ is broken or that the theory is gapless \cite{Gaiotto:2017yup,Komargodski:2017dmc,Komargodski:2017smk,Sulejmanpasic:2018upi,Tanizaki:2018xto}. Both of these are nicely consistent with the limits of 
$m^2\rightarrow \pm \infty$ we discussed above, and the critical to dimer transition of spin-chains (see e.g.~\cite{jullien1983finite,Affleck:1988px,sandvik1999spin,Patil:2018wpt}), provided that the translation symmetry is 
identified with the symmetry\footnote{The label $\C\F$ is there to imply that the transformation is a combination of the $\mathbb Z_2$ flavor 
symmetry subgroup $\F:\Phi\rightarrow i\sigma^2\Phi $ and the $\C$ symmetry. } $\C\F:\Phi\rightarrow i\sigma^2 \Phi^*$, where $\sigma^2$ 
is the standard Pauli matrix. 

It is natural to conjecture that the phase transition between the $m^2\rightarrow \infty $, $\C$-broken phase to the $m^2\rightarrow -\infty$, $O(3)$ 
NLS phase is of the same nature as the phase transition between the dimerized phase of spin chains and the critical phase described by the 
$SU(2)_1$ Wess-Zumino-Witten theory. One argument for this is that the $SU(2)_1$ WZW theory can be deformed to the $O(3)$ NLS 
model at $\theta=\pi$, but only with the use of irrelevant deformations,  indicating that the $O(3)$ NLS model at $\theta=\pi$ would like to 
flow to the WZW theory. 

The picture above is compelling and largely a matter of textbooks by now (see, e.g., \cite{Altland:2006si}). It does, however, rely on many assumptions, such as 
robustness of the critical WZW phase under large irrelevant deformations, the validity of the large $S$ limit down to $S=1/2$, etc. The development of a 
suitable formulation of a lattice model which can be simulated at $\theta \neq 0$, allows us for the first time to provide reliable ab initio data 
for the model \eqref{eq:model}, and to confirm the above picture\footnote{Some numerical experiments on NLS models were performed in 
\cite{Bietenholz:1995zk,Azcoiti:2003vv,Alles:2007br,Azcoiti:2012ws,deForcrand:2012se}, but as opposed to our approach their applicability is limited, and not easily generalizable to other interesting cases.}.

To test whether the transition from the $m^2\rightarrow \infty$, $\C$-broken phase to the $m^2\rightarrow -\infty$ phase is of the same nature as 
the dimer to critical spin chain transition, we must discuss its universal properties. The relevant critical phase of the spin-chains is the $SU(2)_1$ WZW phase \cite{Affleck:1987ch}, which was checked by  numerous simulations \cite{Ziman:1987zz,starykh1997quantum,Tang:2011zz,patil2017indicators,Patil:2018wpt} and is consistent with the $S=1/2$ integrable 
Heisenberg model \cite{Affleck:1988px}. The $SU(2)_1$ WZW model has  no relevant couplings preserving the symmetries of the spin-chain \cite{Affleck:1987ch}, and 
thus the gapless phase is (believed to be) a fairly robust phase. A potential instability of the WZW phase lies in a marginal operator \cite{Affleck:1988px}, with a coupling $g_m$, which is either marginally 
relevant or marginally irrelevant, depending on the sign, and it is this coupling that drives the transition from the WZW to the $\C$-broken phase. When $\theta\ne \pi$, a relevant operator with scaling dimension $x=1/2$ and coupling $g_r\propto (\theta-\pi)$ will be present in the effective action \cite{Gepner:1986wi,Knizhnik:1984nr,Affleck:1988px}. This operator of course breaks the $\C$ symmetry (and translation symmetry of the spin chain). The RG equation for this coupling is
\be
a\frac{dg_r}{da} \; = \; (2-x)g_r \; = \; \frac{3}{2}g_r\;,
\ee
where $a$ is a sliding cutoff (length) scale. At the transition point between the $\C$-broken and the 
WZW phase, $g_m=0$ and  the spin-Peierls mass-gap opens up $M_{SP} \propto g_r^{2/3}$. At finite volume $L^2$, the singular free energy density\footnote{The free energy density transforms under the RG flow with two parts. 
The first comes from integrating the short-distance degrees of freedom, while the second -- so-called singular or homogeneous -- part is due 
to the scale transformation. It is the second part which is relevant for the critical behavior (see, e.g., \cite{cardy1996scaling}).} must 
be of the form
\be
f \; = \; \frac{1}{L^2} F(g_rL^{3/2})\;.
\ee
This follows from the fact that in 2d the free energy density scales like the inverse correlation length squared, and that 
near $g_r=0$ the dependence on 
$g_r$ must be through the combination $M_{SP}L$. For the susceptibility at $g_r=0$ we thus find 
\be
\chi \; = \; \frac{\partial^2}{\partial g_r^2}f\Big|_{g_r=0} \; \propto \; L\;.
\ee
All of this is valid for $g_m=0$, while logarithmic volume corrections need to be taken into account when $g_m\ne 0$ \cite{Affleck:1988px}. 
This means 
that if we plot $\chi/L$ for different volumes, as we vary $m^2$ in the model \eqref{eq:model} there should be a point $(m^2)_c$
where the curves for different 
volumes intersect, provided that the $1/L$ corrections are sufficiently small\footnote{The definition of the topological susceptibility we use is 
shifted by an overall constant, which is a trivial shift in the dual representation that we employ. }.  

\begin{figure}[t] 
   \centering
   \includegraphics[width=0.52\textwidth]{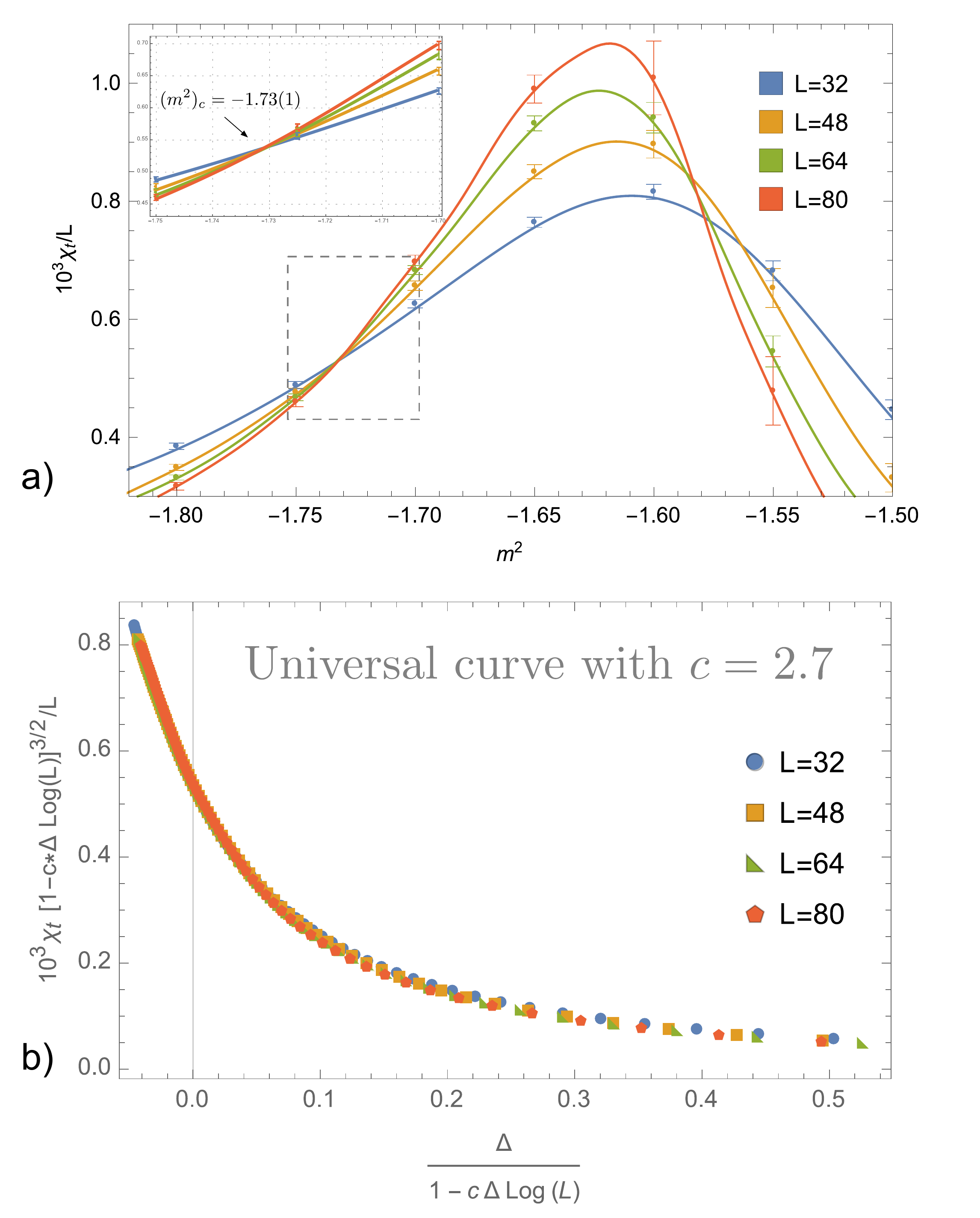} 
   \caption{Reweighted interpolation data for the topological susceptibility $\chi_t$. Fig.~a) shows $\chi_t/L$ for 
   various values of the linear dimension 
   $L$ of the system. All curves intersect well at the point $(m^2)_c=-1.73(1)$, which is consistent with the scaling of the $SU(2)_1$ WZW theory, at 
   marginal coupling $g_m=0$. The inlay show the interpolation data for the dashed region.  Fig.~b) shows that the same data obey the scaling form \eqref{eq:fscaling}, for $c=2.7$. }
   \label{fig:chi_t}
\end{figure}

Fig.~\ref{fig:chi_t}a) shows the numerical data
for four volumes with linear dimension $L=32,48,64$ and $80$, and indeed one can 
clearly see a point where all curves intersect. The simulations were performed for $m^2$ in the interval $[-1.8, -1.5]$, varying $m^2$ in steps of 
$0.05$. This Monte Carlo data was then used to obtain the curves in Fig.~\ref{fig:chi_t}a) using reweighted interpolation. The inlay 
in Fig.~\ref{fig:chi_t}a) shows a zoom into the crossing region for which a separated reweighted interpolation with data from the three indicated points was generated. The four curves intersect at $(m^2)_c=-1.73(1)$ to within the specified accuracy, which gives our estimate of the transition point.

To confirm the nature of the phase transition, we need to derive the scaling form of the topological susceptibility in the presence of a nonzero 
coupling $g_m$. The RG equations for $g_m$ and $g_r$ are \cite{Affleck:1988px}
\begin{align}
&a\frac{dg_m}{da}=\pi b_m g_m^2 \; , &a\frac{dg_r}{da}=\left(\frac{3}{2}+2\pi b_r g_m\right)g_r \; ,
\end{align}
where the sign of $g_m$ is chosen such that $g_m$ is marginally relevant when positive\footnote{Note that this is the opposite convention of that in Ref.~\cite{Affleck:1988px}.}. The constants $b_m$ and $b_r$ are determined by the 3-point functions \cite{Affleck:1988px}, and depend on normalization of the 2-point functions. Indeed, in the above RG equations we can always eliminate either $b_r$ or $b_m$ by redefining $g_m$. One can show that the free energy density at finite volume must be of the form
\be
f=\frac{1}{L^2}\,F\!\left(\frac{\pi b_m g_m}{1-\pi b_m g_m \log(L)},\frac{g_r L^{3/2}}{(g_m)^{{\frac{2b_r}{b_m}}}}\right)\;.
\label{eq:fscaling}
\ee
This result requires some discussion: Under an RG flow the UV cutoff changes as $a\rightarrow a'>a$, while the  linear dimension $L$ shrinks to $\frac{a}{a'}L$, so that $L$ can be 
thought of as changing under the RG flow as the correlation length or inverse mass gap. The overall factor of $\frac{1}{L^2}$ above accounts for the RG flow of the singular free energy density, so that the $F$-function must be constant under the RG flow. For the bare couplings $g_m>0,g_r=0$, an exponentially small mass-gap opens 
$\propto \exp\left(-(\pi b_m g_m)^{-1}\right)$, so the universal function $F$ must depend on the combination 
$\exp\left(-(\pi b_m g_m)^{-1}\right)L$. The first argument of $F$ in \eqref{eq:fscaling} is just the reciprocal of the logarithm of this combination. When $g_r\ne 0$ it is also straightforward to check that the 2nd argument in (\ref{eq:fscaling}) is 
also RG invariant. The same is true for $g_m<0$, so that \eqref{eq:fscaling} holds as long as $g_m, g_r$ are sufficiently small.

Taking the second derivative of \ref{eq:fscaling} with respect to $g_r$ we find
\be
\chi_t= \frac{L}{(1-c\Delta\log L)^{\frac{4b_r}{b_m}}} \, X\!\left(\frac{\Delta}{1-\Delta c \log(L)}\right) \; ,
\label{eq:chiscaling}
\ee
where we set $g_m\propto \Delta=m^2-(m^2)_c$, and introduced an undetermined coefficient $c$, and where $X$ is some universal function.

We already remarked that $g_r\ne 0$ corresponds to a deviation of $\theta$ away from $\pi$, which induces a spin-Peierls transition, such 
that $\frac{2b_r}{b_m}=\frac{3}{4}$ \cite{Affleck:1988px}, and 
the exponent in the pre-factor of (\ref{eq:chiscaling}) is fixed. Fig.~\ref{fig:chi_t}b) shows that the data indeed nicely 
follow the scaling form (\ref{eq:chiscaling}) for a choice of $c=2.7$.

\begin{figure}[t] 
   \centering
   \includegraphics[width=.50\textwidth]{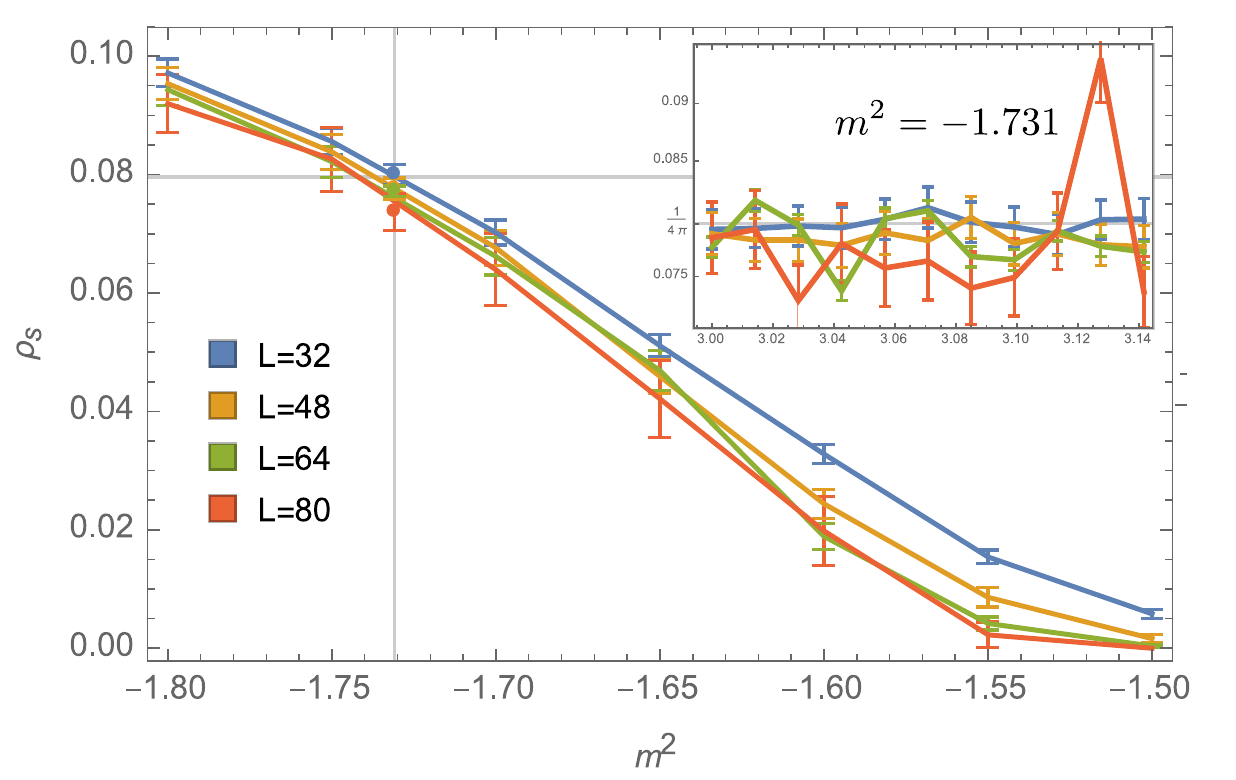} 
   \caption{Spin stiffness as a function of $m^2$. At the transition point $m^2=(m^2)_c = - 1.731$ (marked with a vertical line), 
   where the marginal coupling $g_m=0$ is expected to be zero, it 
   shows good agreement with the universal value $\rho_s=\frac{1}{4\pi}$ (horizontal line). The inlay shows the stiffness at the transition point  for 
   different values of $\theta$ between 3.0 and $\pi$. }
   \label{fig:stiffness}
\end{figure}

As an additional check, in Fig.~\ref{fig:stiffness} we show results of a calculation of the spin stiffness $\rho_s$, which measures the response of a 
system to a constant spatial gauge field $A_1$ for a $U(1)$ subgroup of the $SO(3)$ symmetry, i.e.,
\be
\rho_s\equiv\frac{1}{L^2}\frac{\partial^2 \log Z}{\partial A_1^2}\;.
\ee
The $SU(2)_1$ WZW theory has a description in terms of a compact scalar field $\phi(x)\sim \phi(x)+2\pi$, with Lagrangian\footnote{The model has only a manifest $U(1)$ symmetry, but in fact the symmetry is $SU(2)\times SU(2)$ (see, e.g., \cite{Polchinski:1998rq}). }
\be\label{eq:compact_scalar}
\mathcal L=\frac{1}{4\pi} (\partial_\mu\phi)^2\;.
\ee
The corresponding spin stiffness can be explicitly calculated \cite{SM}
and is given by\footnote{Note that this result is slightly subtle, as the naive expectation 
is that $\rho_s$ is the same as $2$ times the coefficient of the kinetic term in \eqref{eq:compact_scalar}, but this is not the case here (see, e.g.,  
\cite{prokof2000two}).} $\rho_s=\frac{1}{4\pi}$.

In Fig.~\ref{fig:stiffness} we plot the stiffness for various volumes, and indicate the phase transition point (vertical line), as well as the stiffness 
$\rho_s=\frac{1}{4\pi}$ for the $SU(2)_1$ WZW model (horizontal line). As can be seen, exactly at the transition point the stiffness for all volumes
is very close to the expected 
value. We have also computed 
the stiffness at the critical point for values of $\theta$ away from $\pi$ and show the corresponding results in the inlay of Fig~\ref{fig:stiffness}. 

\vspace{0.3cm}

\noindent{\bf Conclusion and future work: } We have presented Monte-Carlo simulations of the lattice 2-flavor $U(1)$ 
gauge-Higgs QFT model, with a 
topological angle $\theta$. We were mostly interested in the value $\theta=\pi$, for which the model is supposed to be an effective description of a 
half-integral spin chain. Such spin-chains with a full $SO(3)$ spin symmetry can be in two phases: in the dimerized phase with a two-fold 
degenerate gapped ground state, and in the critical $SU(2)_1$ WZW phase. We have shown that the lattice discretization we proposed in 
\cite{Gattringer:2018dlw,Sulejmanpasic:2019ytl}, 
which has the correct symmetries and anomalies, gives rise to ab initio results consistent with the expected WZW/dimerized transition. 

This not only complements decades of research on anti-ferromagnetic spin-chains and their connections to QFTs with $\theta$-terms, but also shows the potential of the novel lattice formulation of abelian gauge theories \cite{Gattringer:2018dlw,Sulejmanpasic:2019ytl}, which have applications not only to other interesting 2d models like the asymptotically free $CP^{N-1}$ models, and flag-manifold sigma models related to $SU(N)$ spin chains \cite{Bykov:2011ai,Lajko:2017wif,Tanizaki:2018xto,Ohmori:2018qza}, but also for $U(1)$ gauge theories in higher dimensions. Such formulations allow an enhanced control over monopoles in abelian gauge theories, which are relevant for higher-dimensional spin systems (e.g., for deconfined criticality of anti-ferromagnets in 2 spatial dimensions \cite{Senthil:2004aza}) and the lattice theory of electromagnetism, where monopoles were thought to be an unavoidable curse on the lattice.
\vspace{.3cm}

\noindent{\bf Acknowledgments: } We would like to thank Balt van Reese, Anders Sandvik and Yuya Tanizaki for discussions and comments. TS is supported by the Royal Society of London URF. This work was also supported  by the Austrian Science Fund FWF, grant I 2886-N27.

\bibliographystyle{utphys}
\bibliography{bibliography}

\section{Supplementary Materials}
\begin{appendix}
\section{Lattice formulation, duality and simulation details}

Using the lattice gauge action from Eq.~(2) of the letter and the standard discretization of bosonic matter, i.e., the Higgs fields couple 
to $U(1)$-valued link variables
$U_\mu(x) = e^{iA_{x,\mu}}$, the lattice-discretized partition function for the $U(1)$ gauge-Higgs model we consider reads
\begin{equation}
Z \; = \; \int \! D[A] \, D[\Phi] \; B_G[A;\theta] \, e^{\, - \, S_m[\Phi,A]} \; ,
\label{Zdef}
\end{equation}
where $B_G[A;\theta]$ is the gauge field Boltzmann factor
\begin{equation}
B_G[A;\theta] \; = \; \prod_{x\in\Lambda} \sum_{n_x\in\mathbb{Z}} 
e^{\, - \, \frac{\beta}{2}\left(F_x+2\pi n_x\right)^2 \, - \, i \theta n_x} \; ,
\label{B_G}
\end{equation}
with $F_x = A_{x+\hat{1},2} - A_{x,2} - A_{x+\hat{2},1} + A_{x,1}$. The matter action $S_M[\Phi,A]$ is given by
\begin{eqnarray}
S_H[\Phi,A] & \; = \; & \sum_{x}\Big[ M \Phi_x^\dagger \Phi_x  + 
\lambda \big(\Phi_x^\dagger \Phi_x\big)^2 
\nonumber \\
& & \quad - \sum_{\mu=1}^{2} 
\big(\Phi_x^\dagger e^{iA_{x,\mu}} \Phi_{x + \hat \mu} \, + \, c.c.\big) \Big] \; ,
\label{S_Higgs}
\end{eqnarray}
with the mass parameter $M$ defined as $M = 4 + m^2$, where $m$ is the bare mass. The variable $x$ runs over all sites
of an $L_1 \times L_2$ square lattice where all fields obey periodic boundary conditions. The measures in the partition function 
(\ref{Zdef}) are the usual product measures over all degrees of freedom on the sites and links of the lattice.  

With the help of Poisson resummation discussed in Eq.~(3) of the letter we can linearize the dependence of the Boltzmann factor (\ref{B_G}) 
on $F_x$ and after an expansion of the nearest neighbor term Boltzmann factors of the Higgs field integrate out the gauge fields
$A_{x,\mu}$ and the matter degrees of freedom $\Phi_x$ (see \cite{Gattringer:2018dlw,Sulejmanpasic:2019ytl} for more details). 

The result is an exact rewriting of the partition sum (1) in terms of flux variables $j_{x,\mu}, k_{x,\mu} \in \mathds{Z}$ 
which describe the two matter field components and the plaquette based 
variables $m_x \in \mathds{Z}$ for the gauge degrees of freedom. In this dual form the partition sum is a sum over the 
configurations of the new variables,
\begin{eqnarray}
\hspace*{-0mm} && Z  =   \sum_{\{m\}}  W_G[m]  \; \sum_{\{j,k\}} W_H[j,k] \;
\prod_x  \delta\!\left(\!\vec{\nabla}\cdot \vec{j}_x\! \right) \delta\!\left(\!\vec{\nabla}\cdot \vec{k}_x\! \right) 
\label{dualZ} \\
\hspace*{-0mm} &&
\times \, \prod_x \!  
\delta\!\left(j_{x,1}\!+\!k_{x,1}\!+\!p_{x}\!-\!m_{x-\hat{2}}\right)  \delta\!\left(j_{x,2}\!+\!k_{x,2}\!-\!m_{x}\!+\!m_{x-\hat{1}}\right)\! .
\nonumber
\end{eqnarray}
In (\ref{dualZ}) the sums over the configurations of the new variables are defined as 
$\sum_{\{m\}} \equiv  \prod_x \sum_{m_x \in \mathds{Z}}$ and 
$\sum_{\{j,k\}} \equiv \prod_{x,\mu} \sum_{j_{x,\mu}\in\mathds {Z}} \sum_{k_{x,\mu}\in\mathds{Z}}$. 
The weight factor $W_G[m]$ for the plaquette occupation numbers $m_x \in \mathds{Z}$ is given by 
($V = L_1 L_2$ denotes the number of sites) 
\begin{equation}
W_G[m] \; = \; \left( 2\pi \beta \right)^{-\frac{V}{2}} 
 e^{\, - \frac{1}{2 \beta} \sum_x\big(m_x + \frac{\theta}{2\pi}\big)^2} \; .
\label{W_G}
\end{equation}
The weight factor $W_H[j,k]$ for the scalar fields is a sum over link-based auxiliary variables 
$a_{x,\mu}, b_{x,\mu} \in \mathds{N}_0$ with  $\sum_{\{a,b\}} \equiv 
\prod_{x,\mu} \sum_{a_{x,\mu}\in\mathds {N}_0} \sum_{b_{x,\mu}\in\mathds{N}_0}$,
\begin{eqnarray}
&& W_H[j,k] =  \sum_{\{a,b\}} \, \prod_x \! I\left(f_x,g_x\right) 
\nonumber \\
&& \hspace{10mm} \times \, \prod_{x,\mu} \, 
\frac{1}{\left(|j_{x,\mu}|\!+\!a_{x,\mu}\right)! \ a_{x,\mu}!} \, 
\frac{1}{\left(|k_{x,\mu}|\!+\!b_{x,\mu}\right)! \ b_{x,\mu}!} \; ,
\nonumber \\
&& 
I\!\left(f_x,g_x\right)  =  \int_{0}^{\infty} \!\!\!\!\! dr \!\!\!  \int_{0}^{\infty} \!\!\!\!\! ds \, r^{\,f_x+1} \, s^{\,g_x+1} \, 
e^{\, - \, M (r^2 + s^2) \, - \, \lambda (r^2 +  s^2)^2 } ,
\nonumber \\
&& 
f_x = \sum_{\mu}
\left[ |j_{x,\mu}| + |j_{x-\hat{\mu},\mu}| + 2\left( a_{x,\mu} + a_{x-\hat{\mu},\mu} \right) \right] \; ,
\nonumber \\
&& 
g_x =\ \sum_{\mu}
\left[ |k_{x,\mu}| + |k_{x-\hat{\mu},\mu}| + 2\left( b_{x,\mu} + b_{x-\hat{\mu},\mu} \right) \right] \; .
\label{W_H}
\end{eqnarray}
Obviously the weight factors (\ref{W_G}) and (\ref{W_H}) are real and positive for all  
$\theta$, such that the sign problem is solved. 

The dual variables $j_{x,\mu}, k_{x,\mu}$ and $m_x$ obey constraints that are written as products of
Kronecker deltas in (\ref{dualZ}), where we use the notation $\delta(n) = \delta_{n,0}$. These constraints 
enforce vanishing divergence $\vec{\nabla}\cdot \vec{j}_x = \sum_\mu [ j_{x + \hat \mu,\mu} - j_{x,\mu}]$ at all sites $x$
for both, the $j$ and the $k$ variables, which implies that they must form closed loops of flux. 
The second set of constraints involves all three types of the dual variables and enforces that for all links the combined oriented flux 
of $j$, $k$ and the plaquette variables $m_x$ on the plaquettes that contain the link must add up to 0. 
  
A suitable Monte Carlo update of the dual form (\ref{dualZ}) must take into account the constraints. In our simulation this is 
implemented by a mix of updates that ensure ergodicity and a reasonably fast decorrelation of the configurations of the dual variables.
More specifically we change plaquette occupation numbers $m_x$ by $\pm 1$ and simultaneously change the flux of $j$ or $k$
around that plaquette. This is combined with worms \cite{Prokofev:2001ddj} for doubly occupied loops where we jointly update  
$j$- and $k$- fluxes of opposite sign, as well as surface worms \cite{Mercado:2013yta} that jointly update flux and plaquette variables. 
Furthermore we include a global change of the plaquette occupation numbers by proposing to
 change all $m_x$ by the same value $\pm 1$, and finally also make use of the $\mathds{Z}_2$ symmetries of our system by performing 
 charge conjugation and flavor swapping updates on our configurations.

We consider various lattice sizes $V = L \times L$ with $L$ ranging from $L = 32$ up to $L = 112$. The couplings
$\lambda$ and $\beta$  are kept fixed at $\lambda = 0.5$  and $\beta = 3.0$ 
for this work and we study the system as a function of the mass parameter 
$M = 4 + m^2$. Typically we use statistical sample sizes of $10^4$ to $10^7$ configurations, which are separated by 
10 to 100 decorrelation sweeps. The initial equilibration is performed with $5 \times 10^5$ sweeps. The error bars are estimated by Jackknife, combined with binning to account for autocorrelations.

\section{Stiffness in compact scalar theory}

The compact scalar Lagrangian in Euclidean space is given by
\be
\mathcal L=\frac{J}{2}(\partial_\mu\phi)^2\;,
\ee
where compactness implies the identification $\phi\sim \phi+2\pi$. For computing the spin-stiffness 
we couple the model to a constant gauge field  by replacing $\partial_1\phi \rightarrow \partial_1\phi+A_1$. Expanding the action we find
\be
\mathcal S=\int d^2x\;\Big[\frac{J}{2}\left(\partial_\mu\phi\right)^2+J\partial_1\phi A_1+\frac{J}{2}A_1^2\Big]\;.
\ee
We are interested in the theory on a torus with $L_1,L_2$ being the lengths of the two cycles.
Now we write 
\be
\phi(x_1,x_2)=\frac{2\pi Q x_1}{L_1}+\tilde\phi(x_1,x_2)\;,
\ee
where $\tilde \phi(x_1,x_2)$ is periodic, and $Q=\int dx^1\partial_1\phi$ is the winding number along the cycle $1$. The action turns into
\begin{equation}
\mathcal S=\frac{J}{2}\int d^2x\; (\partial_\mu\tilde\phi)^2+\frac{JL_2}{2L_1}\left({2\pi Q}+A_1L_1\right)^2 .
\end{equation}
Obviously the  field $\tilde\phi$ decouples from the rest, such that we can consider the partition function
\be\label{eq:Z}
Z=\sum_{Q\in \mathbb Z}e^{-\frac{JL_2}{2L_1}\left({2\pi Q}+A_1L_1\right)^2}\;.
\ee
Note that the sum above can be expressed in terms of the Jacobi theta function $\vartheta_3$, but we will not need this form for what follows.

The stiffness is given by
\be
\rho_s=-\frac{1}{L_1L_2}\frac{\partial^2\log Z}{\partial A_1^2}\Big|_{A_1=0}=-(2\pi)^2J^2 \frac{L_2}{L_1}\avg{Q_1^2}+J\;,
\ee
where the average is taken with the partition function \eqref{eq:Z} at $A_1=0$. 

\begin{figure}[tb] 
   \centering
   \includegraphics[width=.44\textwidth]{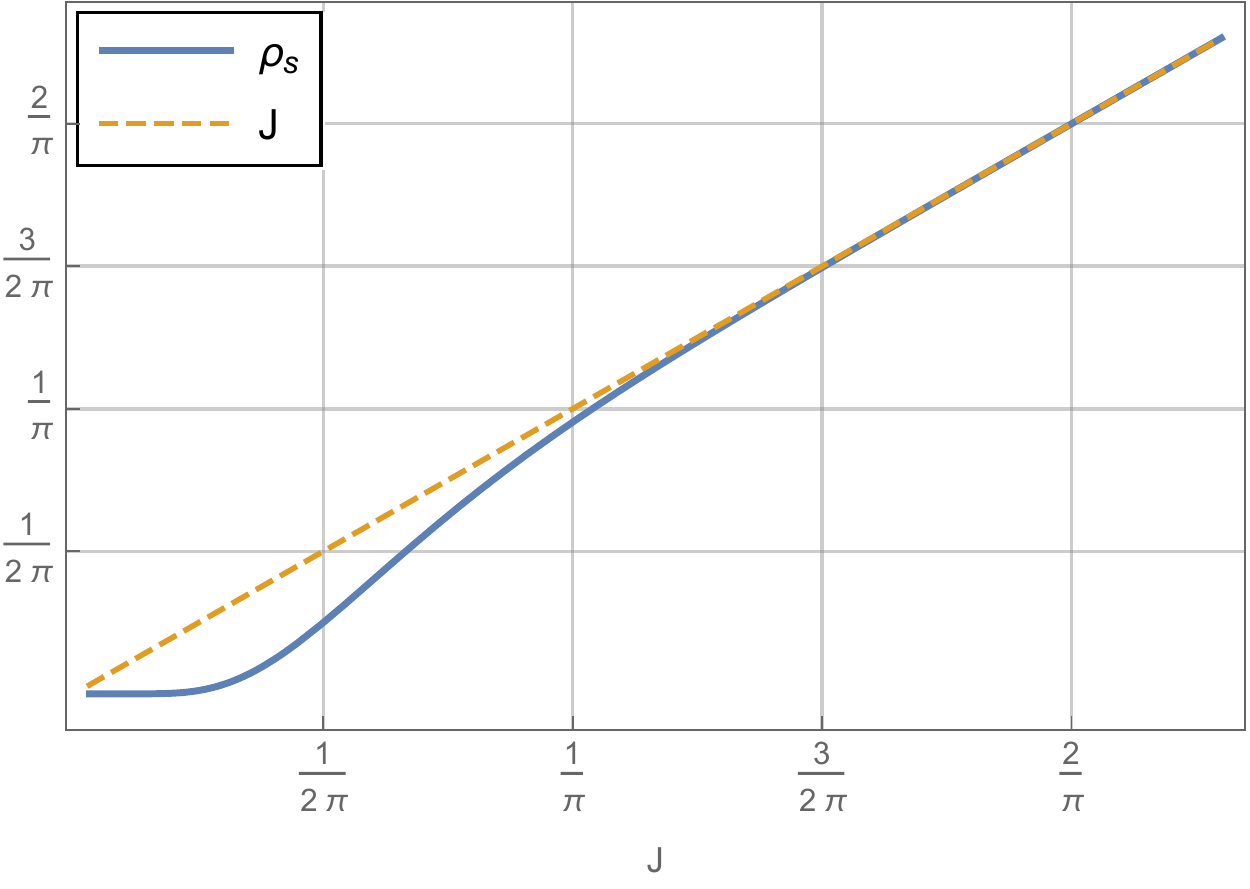} 
   \caption{Plot of the stiffness from the full partition sum (\ref{eq:Z}), and comparison with the result 
   $\rho_s\approx J$ valid for large $\frac{JL_2}{L_1}$.}
   \label{fig:fig1}
\end{figure}

If the combination $\frac{JL_2}{L_1}$ is sufficiently large, all the values of $Q$ away from zero are exponentially suppressed in (\ref{eq:Z}), 
and we immediately find $\rho_s=J$. This is the usual result, which is valid to significant accuracy, for the usual BKT transition for $L_1=L_2$, that happens at $J=\frac{2}{\pi}$ \cite{prokof2000two}. Here, however, we are interested in the point $J=\frac{1}{2\pi}$, which is not sufficiently large to make the above approximation. The plot in Fig.~\ref{fig:fig1} shows the comparison of the stiffness result from the full partition sum (\ref{eq:Z}) at 
$L_1=L_2$ with the function $J$ to illustrate this.

On the other hand the partition function \eqref{eq:Z} is periodic with respect to the magnetic flux $A_1L_1$ with a period $2\pi$, as can be seen by inspection. The Fourier expansion thus is given by
\be
Z=\sum_{w\in \mathbb Z} \sqrt{\frac{L_1}{2\pi JL_2}}e^{-\frac{w^2L_1}{2JL_2}+i w A_1L_1}\;.
\ee
The stiffness therefore reads
\be\label{eq:Zdual}
\rho_s=\frac{L_1}{L_2}\avg{w^2}\;,
\ee
where again the average is taken at $A_1=0$. Now if $\frac{JL_2}{L_1}$ is sufficiently small, we find that $\rho_s=O(e^{-\frac{L_1}{2JL_2}})$.

We are interested in the case $L_1=L_2=L$ and $J=\frac{1}{2\pi}$, which is between the two extremes of small and large $\frac{JL_2}{L_1}$. However, in this case it is easy to see that $\avg{w^2}=\avg{Q^2}$. Now identifying \eqref{eq:Z} and \eqref{eq:Zdual}, we find
$\avg {Q^2}=\avg{w^2}=\frac{1}{4\pi}$, and obtain
\be
\rho_s=\frac{1}{4\pi}\;.
\ee

\end{appendix}

\bibliographystyle{apsrev4-1} 

\end{document}